# ASTRONOMICAL TENT OBSERVATORIES, RELICS OF A BYGONE ERA

**Richard de Grijs**
*School of Mathematical and Physical Sciences, Macquarie University,*
*Balaclava Road, Sydney, NSW 2109, Australia*
Email: richard.de-grijs@mq.edu.au

**Abstract:** Between the mid-eighteenth and mid-nineteenth centuries, long-haul oceanic voyages of exploration and discovery routinely carried astronomical tent observatories to support land-based longitude determinations using heavy and cumbersome astronomical regulators and transit telescopes. Following James Cook's deployment of a pilot tent observatory on his first voyage to the Pacific in 1768–1771, the tent design was altered by William Bayly for more convenient use on Cook's second and third voyages to the Pacific. Bayly's design became the standard structure of tent observatories assigned to shipboard astronomers during the Age of Sail. By the middle of the nineteenth century, a subtle shift in focus had occurred, with tent observatories now being deployed to observe specific celestial events (such as the 1882 Venus transit or a variety of eclipses), while longitude determinations increasingly relied on the novel, compact and improved box chronometers of the day. A further shift in the application of tent observatories occurred towards the end of the nineteenth century, when astronomical applications largely gave way to a renewed focus on meteorological measurements.

**Keywords:** tent observatories, James Cook, Venus transits, territorial surveys, longitude determination

## 1 SHIP OR SHORE?

By the end of the 'Age of Sail', which lasted roughly from the mid-fifteenth to the mid-nineteenth centuries, it had become common for long-haul oceanic voyages of discovery and exploration to enlist professional astronomers among ships' crews. Their services were indispensible for accurate geographic position determinations once the expedition had strayed beyond familiar waters.

Determination of one's latitude was straightforward and involved measurements of either the Sun or known bright stars as they crossed the local meridian. Longitude determination, on the other hand, required access to an accurate timepiece that reflected the time at the ship's home port (or at a reference meridian). Alternatively, longitudes could be determined by means of extensive, cumbersome calculations based on lunar distance measurements (e.g., de Grijs, 2020), combined with up-to-date nautical almanacs that included tables of lunar distances and their timings for one's reference meridian.

Accurate, compact (box) chronometers eventually become more commonplace on voyages of discovery by the mid-nineteenth century. Until then, however, expeditions' time measurements relied on a combination of early marine chronometers and astronomical lunar distance observations. Early marine timepieces often gained or lost time over the course of a voyage, and so lunar distance observations were essential for calibration purposes, that is, lunar distances were used to check the 'going rates' of shipboard clocks.

When in port or close to shore, such checks were often preferentially done on land and based on the most accurate time measurements available, those obtained from the type of long-case regulators carried on most long-haul voyages. Such 'grandfather' clocks often took several days to set up and 'settle', however (e.g., Davies, 2022: Pt 2, Ch. 10), and they could only be used appropriately on land. In addition to those rather awkward clocks, the astronomer's arsenal of instruments required for accurate time measurements included one or more sextants, a universal theodolite, chronometers and nautical almanacs, and often also a smaller journeyman clock. To protect these instruments from the vagaries of the weather (and sometimes from the local population), from the mid-eighteenth century shipboard astronomers would often set up canvas tent observatories on shore.

Tent observatories were most often set up on a wooden frame. The instruments were generally placed on wooden casks, originally used to transport food, water, ale, wine or rum, which were filled with heavy, wet sand as ballast for stability (e.g., Orchiston, 2004; 2017). The literature on astronomical tent observatories, their design and their use is scattered and incomplete. Very few articles cover any specifics, and if they do, they tend to focus on narrowly constrained niche areas (e.g., Hawkins, 1979; Orchiston and Rowe, 2021). Here, I intend to provide a more comprehensive account covering the era of the astronomical tent observatory.

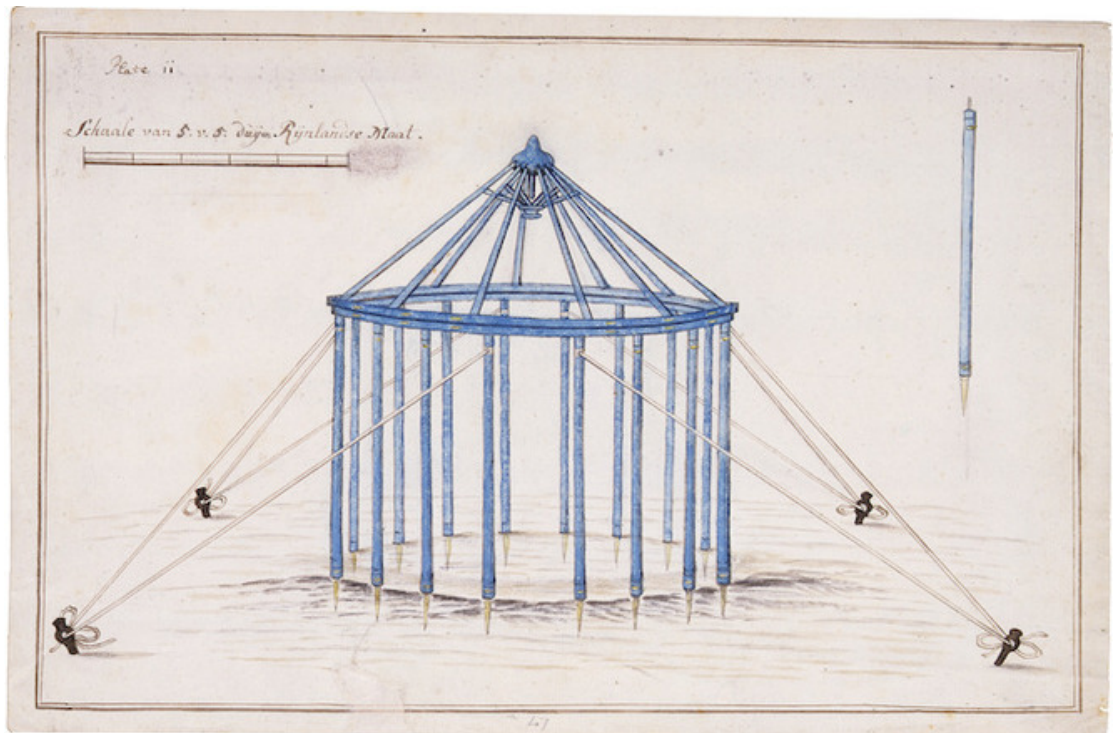

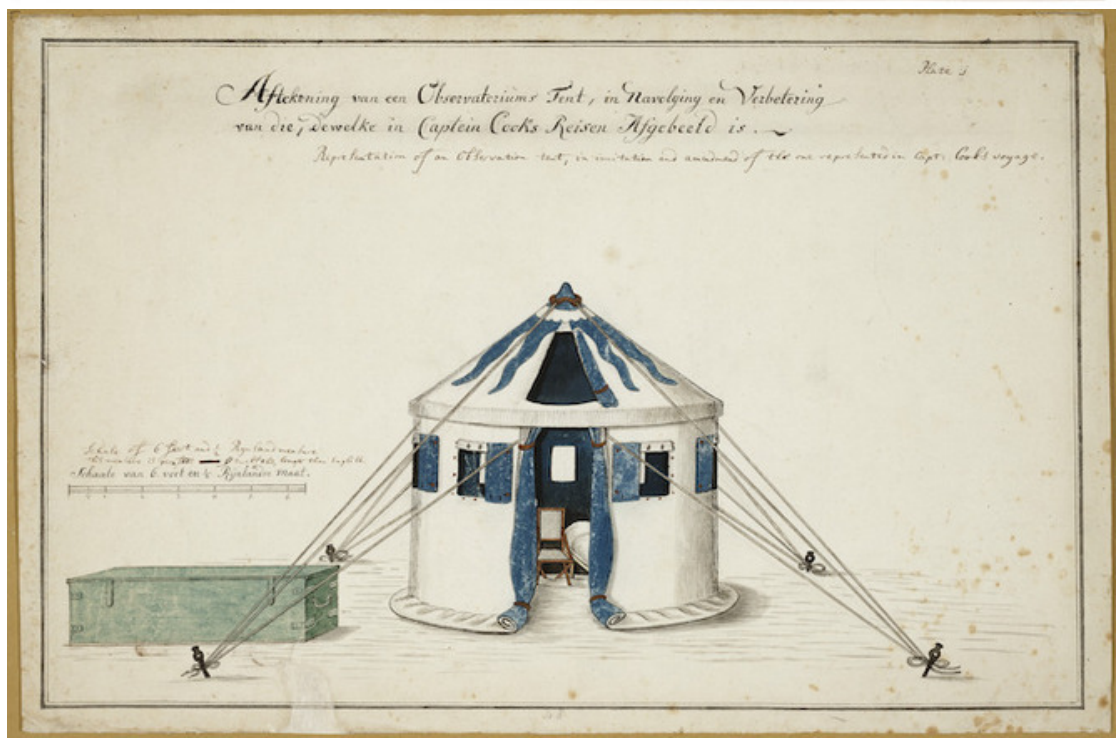

**Figure 1**: Eighteenth-century tent observatory of the type routinely supplied to shipboard astronomers on long-distance oceanic voyages of exploration. Credits: (*top*) Artist unknown. Alexander Turnbull Library, Wellington, New Zealand (Ref. B-091-008, /records/23199229). (*bottom*) Artist unknown. Aftekening van een observatoriums tent in navolging en verbetering van die, dewelke in Captain Cook's reisen afgebeeld is. Plate 1, ca. 1780? Alexander Turnbull Library, Wellington, New Zealand (Ref. B-091-007, /records/22320145) (both images reproduced with permission).

Having established stable shore-based observatories allowed the astronomers to obtain more accurate measurements than could be obtained on rocking and pitching ships. At sea, they would obtain observations to determine their geographic coordinates, as well as magnetic and meteorological observations. On land, they would additionally record tidal observations, whereas the stable operation of a pendulum clock would also allow them to obtain gravity measurements (e.g., Morrison and Barko, 2009; Bosloper, 2010, 2017; de Grijs, 2022).

As we will see below, however, tent observatories were not in use until James Cook's (1728–1779) first voyage to the Pacific, New Zealand and Australia of 1768–1771. They were, in fact, only used for a relatively short period, until the development of reliable, compact maritime timepieces had made the use of tent observatories on maritime voyages of exploration redundant by the end of the eighteenth century. One of the last major sea-bound explorations carrying a tent observatory was Matthew Flinders' 1801–1802 circumnavigation of Australia on H.M.S. *Investigator*. Matthew Flinders (1774–1814), and increasingly his brother Samuel Ward Flinders (1782–1834), used a Ramsden universal theodolite, a Hadley sextant and an Earnshaw astronomical regulator to determine their longitudes using the lunar distance method. Pitching a tent observatory on shore, they then used their lunar distances to check the 'going rate' of their clocks (Kenneth, 2017).

## 2 CAPTAIN COOK'S TENT OBSERVATORIES

In the late 1760s, the Royal Society of London actively encouraged observations of the 1769 transit of Venus across the face of the Sun, funding several expeditions to do so. Venus transits, particularly when combining observations from multiple, geographically distinct vantage points, offer a unique opportunity to determine the distance from the Earth to the Sun, and hence to determine the size of the solar system as a whole. Cook's first voyage to the Pacific on H.M. Bark *Endeavour* was one such expedition. It was reportedly the first scientific expedition equipped with an experimental prototype tent observatory. Under the supervision of Nevil Maskelyne (1732–1811), Britain's fifth Astronomer Royal, and designed by John Smeaton (1724–1792), a civil engineer, it was a heavy and rather cumbersome canvas structure (see Figure 1).

The *Endeavour* arrived in Matavai Bay, on Tahiti's northern coast, on 13 April 1769, well in time to prepare for the transit observations of 3 June of that year. Cook selected Point Venus, a narrow peninsula known locally as Te Auroa, as the expedition's base. The protruding landmass represented an ideal observatory site; it could be accessed easily from the ocean and it was readily fortified as well. The *Endeavour*'s crew proceeded to build a banked, ditched and pallisaded enclosure containing the expedition's tents and equipment to "… protect the observers and the instruments from the natives …" (Badger, 1969: 37),[1] which became known as Fort Venus. Its defences were bolstered by some of the cannons and swivel guns from the *Endeavour*.

Cook set up a tent containing a clock with a grid-iron pendulum, established on a wooden frame firmly fixed in the ground:

> The astronomical clock, made by Shelton and furnished with a gridiron pendulum, was set up in the middle of one end of a large tent, in a frame of wood made for the purpose at Greenwich, fixed firm and as low in the ground as the door of the clock-case would admit, and to prevent it being disturbed by any accident, another framing of wood was made round this, at the distance of one foot [30.5 cm] from it. (Green and Cook, 1771: 397).

The pendulum's length was adjusted to the same length as that at Greenwich to allow for gravity observations (e.g., Bosloper 2010, 2017; de Grijs, 2022). Green and Cook (1771: 398) later described the camp's design, pointing out that facing the tent containing the long-case regulator "… stood the observatory, in which were set up the journeyman clock and astronomical quadrant: this last, made by Mr. [John] Bird [1709–1776], of one foot radius, stood upon the head of a large cask fixed firm[ly] in the ground, and well filled with wet heavy sand". The observatory also contained three pivoted, reflecting telescopes made by James Short (1710–1768; Turnbull, 2004).

Cook was apparently lulled into a false sense of security: "I now thought myself perfectly secure from anything that these [native] people could attempt …" (cited by Captain Cook Birthplace Museum, n.d.). However,

> … when Mr. [Charles] Green and I went to set up the Quad.[t] it was not to be found, it had never been taken out of the Packing case … and the whole was pretty heavy, so that it was a matter of astonishment to us all how it could be taken away, as a Centinal [*sic*; sentinel] stood the whole night within 5 yards [4.6 m] of the door of the Tent where it was put … (*Ibid.*)

The quadrant was eventually recovered—and repaired—well before it was required for the Venus transit observations.

On Cook's second and third voyages to the Pacific of 1772–1775 and 1776–1779, respectively, the heavy and cumbersome Smeaton observatory design was discarded. Instead, Cook's astronomer William Bayly (1737/8–1810) designed "… undoubtedly one of the most convenient portable Observatories that had yet been made", equipped with hinges that allowed "… the roof to open and close like an umbrella." (Wales and Bayly, 1777: vii). Bayly's and Smeaton's observatory designs replaced the less flexible, rigid wooden huts commonly used on earlier voyages.

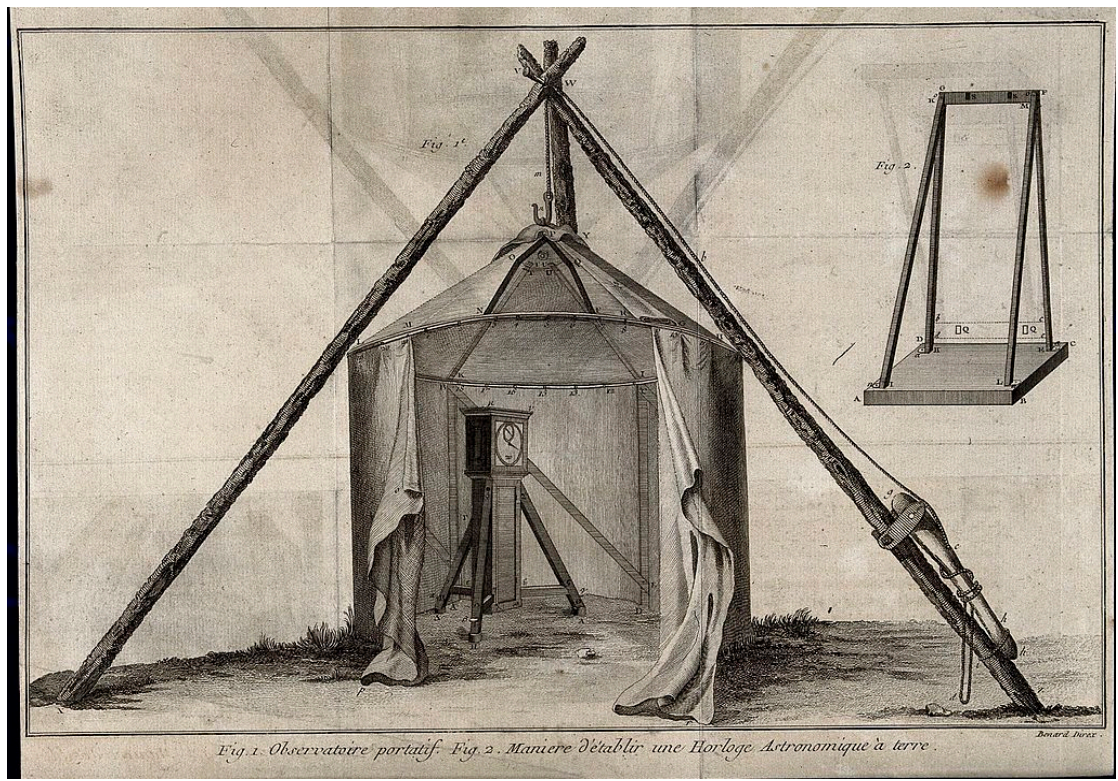

**Figure 2**: Tent observatory of Bayly's design (Wales and Bayly, 1777). Engraving by Benard (possibly after L. J. Goussier). (Wellcome Library, no. 46268i; public domain).

Cook undertook his second voyage to the Pacific and down to Antarctic waters with two ships, H.M.S. *Adventure* (commanded by Captain Tobias Furneaux, 1735–1781) and H.M.S. *Resolution*, under Cook's command. Bayly was assigned to astronomy duties on the *Adventure*, whereas William Wales (1734?–1798) served as the *Resolution*'s astronomer. Both were equipped with identical tent observatories, which "... will be of infinite use to them whenever they may have occasion to make observations on shore ..." (cited by Beaglehole, 1969: 722). Wales and Bayly (1777) provided a detailed description of their observatory's construction:

> The upright sides consist of eight staves, *AB*, *CD*, &c. (see Plate II) [Figure 2] about two inches [5.1 cm] diameter, and five feet and an [*sic*] half [1.68 m] long, which supported a circular ring, 1, 2, 3, 4, &c. to 21. of eight feet [2.44 m] diameter, and the covering, *r*, *q*, 9, 10, &c. to 21, *o*, *p*, of oiled canvass. The staves are of beach-wood [*sic*; beech-wood], armed at the bottom with spikes, to stick into the ground, and at the top with small iron pins, fitted to holes which are made to receive them in the ring. The ring is composed of eight circular arches, of about three feet [91 cm] long, two inches broad, and an inch [2.5 cm] thick, made of beach-wood, and are readily put together, or taken asunder by means of strong iron plates, screwed fast with wood screws to the end of one arch, and by screws and nuts to the end of another, for the purpose of frequent screwing and unscrewing without danger of wearing out the holes, as would be the case with wood screws entering the wood itself. Into the outer edge of this ring are drove small staples, 1, 2, 3, 4, &c. and to the upper edge of the canvass, answerable thereto, are sewed several small hooks, which being hooked into these staples, serve to support the upper edge of the canvass, while its lower edge just reaches to the ground: The two parts of the canvass, 2, 1, *o*, *p*; 9, *q*, *r*, are supposed to be unhooked from the staples 1, 2, 3, 4, and 5, 6, 7, 8 respectively, and thrown back to shew inside the Observatory, and the manner of fixing up the Clock, …: *BE* is a brace of the same sort of wood, screwed fast to the top of the staff *AB*, by a screw at *B*, and to the bottom of the staff *CD* at *E*. These braces, from the top of one staff to the bottom of the next, kept the whole upright circular frame very steady. *FGHIKLMN* is another circular ring, exactly of the same dimensions and construction with the former, on which it rests: To this the roof of the Observatory is screwed by means of ten long screws, which pass through the end of the rafters at *FGHIK*, &c. into iron nuts fixed in this upper ring for that purpose. The rafters *MP*, *RP*, *IU*, *KE*, &c. are attached to the crown-piece *PTU* by hinges, as represented at *T*, and *U*; and the two short rafters *FQ*, *NO*, are attached to the two *RP*, *MP*, also by hinges at *O*, and *Q*: By means of these hinges the roof is made to open or close like an umbrella, and of course, if disengaged from the circular ring *FRH*, &c. will fold together, and may be packed up in a very small compass.

The covering of the roof is of very thick canvas, many times painted, and comes down so far as to hang over at the eves about four inches [10.2 cm]. The crown-piece, *PTU*, is about eight inches [20.3 cm] in diameter, and covered with a circular piece of canvas like that the roof is covered with. An eye-bolt *n*⊙ passes through its center [*sic*], and is fastened on the inside by the nut ⊙. This eye-bolt is intended for the reception of the hook *n*, which is fastened to the cord *mbgcd*, passing over a pulley at *W*, fixed in the top of the pole *VZ*. Towards the bottom of this pole there is fixed a lever *gh*, by means of the clamp *ef*, and its fellow on the opposite side, and the lever, turns on the iron bolt *f*. The cord *mbcd* passes through a hole *c* in the lever, and is drawn right when the end *b* of the lever is turned upwards, and then made fast: Now if the end *b* of the lever be brought down towards *z*, and there fastened by means of the becket [loop of rope], or endless cord *ik*, the roof of the Observatory will be drawn up from off the ring 1, 2, 3, &c. and may be turned round by twisting or unwinding of the cord, until the opening *NOPQE* is towards the sun, or any other object, of which an observation is wanted to be made. When the observation is completed, the lever may be released, and the roof is let down again to rest with its whole weight on the lower ring, as it will then be less liable to be disturbed by the wind: There are also eight small staples on the inside edge of the lower ring 1, 2, 3, &c. and as many small hooks, corresponding to them, on the upper, or that to which the rafters of the roof are fastened. These hooks, when the roof is lowered down, are to be hooked into the staples, and the cord then drawn tight, to prevent, yet farther, the effect of the wind. The opening *N*, *OP*, *QF*, is covered when not in use, by the flap, or roll of spare canvas *QRGS*, which is of the same sort, and painted in the same manner, as that which covers the roof. The whole of this Observatory, except the three poles *WZ*, *WX*, and *WY*, when taken down and packed up properly, is contained in a chest six feet and nine inches [2.06 m] long, and about 20 inches [50.8 cm] square: The poles, which form the tripod, are of about fifteen feet [4.57 m] long, and four inches diameter, may be laid amongst the spare booms of the ship, or if they should be thought too cumbersome there, may be cut out of the woods, or purchased for a trifle at any place where they are wanted. (Wales and Bayly, 1777: viii–x).

Wales and Bayly were both equipped with a similar complement of instrumentation as Cook and Charles Green (1734–1771) had carried on Cook's first voyage. They carried Gregorian reflecting telescopes (manufactured by Bird), astronomical quadrants, sextants, clocks and watches. In addition, the astronomers now had access to two Dollond refracting telescopes, a transit telescope constructed by Bird and four chronometers, which were modeled after John Harrison's (1693–1776) prize-winning H4 'longitude clock' (Orchiston and Howse, 1998). Wales was given a chronometer made by John Arnold (1736–1799) and also Larcum Kendall's (1719–1790) 'K1', an exact copy of Harrison's H4. Bayly was equipped with two Arnold chronometers (Betts, 1993).

The *Resolution* and *Adventure* left Plymouth Sound on 13 July 1772. Bayly's 'observations book' offers an interesting account of the difficulties shipboard astronomers would often encounter in attempting to carry out their duties on foreign shores. We get the distinct impression that setting up their tent observatories and the larger astronomical instruments and pendulum clocks was anything but straightforward (e.g., Baker, 2012). Nevertheless, the scientists exhibited patience and creativity in overcoming most obstacles.

For instance, in July 1772 Bayly screwed his astronomical regulator to a heavy bookcase in the Consul's House at Funchal (Madeira), because "… the book-case stood on a floor that was paved with Bricks & it was full of Books which rendered it very steady" (Bayly, 1772a). Upon arrival at Table Bay by October 1772, he had to overcome additional problems. His initial attempt at weighing down the stand of his astronomical quadrant by filling it with water failed when the water began leaking, and so he resorted to using sand instead. Moreover, on 4 November 1772, they "… Had a strong wind blowing at the SSE which brought great Quantities of sand from the Table mountains which greatly Affected the Instruments by Covering them with sand & shaking them & It was with difficulty we secured the tents from oversetting" (Bayly, 1772b). Both ships left the Cape of Good Hope on 22 November 1772.

On 7 April 1773, the *Adventure* reached Queen Charlotte Sound in New Zealand, where they anchored in Ship Cove near Hippah Island, initially referred to as 'Observatory Island'. Once again, Bayly was forced to creatively solve a number of practical problems before he could proceed with his astronomical observations. Since Bayly did not get along with Furneaux and the other officers on the *Adventure*, the astronomer could only rely on his own servant and an unofficial assistant, "... a good natured Welshman, who would always work if I gave him Brandy ..." (Bayly, 1773a: 207). On 12 April, Bayly recorded in his observations book that

> On the top of this Island saw the only favourable spot I could find any where near the Ship to Observe Equal Altitudes with any propriety, but twas with the greatest difficulty I got the Instruments up, being first obliged to make steps in the Rocks, but as I had only two men sent with me, [they] were only able to cut away the Weeds in order to set up my Tent & make the Aforesaid steps that day. (Bayly, 1773b: 27r).

On 16 April, he "[c]arried [his] Instruments & the Tent Observatory on Shore, together with the iron blocks & [illegible] belonging to the Astronomical Clocks, & it was with the utmost difficulty we got them to the top of the Island" (*ibid*.). They pitched their tent observatory and built a temporary hut for the transit telescope.

On 19 April Bayly set up his astronomical regulator on Hippah Island:

> In one corner of the House (I had built) I sank a hole in the Rock about 12 Inches [30 cm] from Each Side of the House, & about 16 Inches [41 cm] deep; in the bottom of this hole on the Solid Rock, the Block of Iron [to support the clock] was laid as nearly Horizontal as possible, so that the bottom of the Clock-case was about 14 Inches [36 cm] below the floor of the House, & the Clock stood quite independent of every part of it. (*ibid*.)

Eventually, on 21 April he set up the transit telescope, with further adjustments made on 24 April:

> In the morning I moved the Transit Instrument very near the true Meridian by means of the Adjusting Screws, where it Accidentally cut two good marks on the tops of two Hills one to North distant about 1½ Mile [2.4 km]; & the other to the South about 4½ Miles [7.2 km]; the Instrument was constantly kept to these marks during the whole time it remained up, by examining it three or more times a day, & frequently before each observation when in the Day: as was likewise the Level, & the Posts was so well rammed that I never found it [out] more than a bredth [*sic*] of the Wire ... (*ibid*.)

Having become separated from the *Adventure* in heavy fog on 8 February 1773, the *Resolution* eventually arrived at their pre-arranged rendezvous location in Queen Charlotte Sound on 17 May 1773, following a short sojourn at the safe and deep anchorage in Pickersgill Harbour[2] at Dusky Bay (see Figure 3), today known as Dusky Sound. The *Resolution*'s qualified astronomers (Cook and Wales) had not been idle, however. They used their reprieve from the high seas to acquire astronomical observations in Dusky Bay, specifically to determine the site's longitude and latitude.

During Cook's second voyage to the Pacific, the *Resolution* returned twice more to Queen Charlotte Sound, in 1773 and 1774. On all three occasions, Wales diligently recorded their latitude and longitude (see Orchiston, 1997; Orchiston and Howse, 1998). The *Adventure* also returned to Queen Charlotte Sound, but only once, on 1 December 1773, and they missed a rendezvous with the *Resolution* by six days on that occasion. Position determination was also Bayly's principal concern, particularly given the Sound's great potential as a replenishment destination for future Pacific voyages. Bayly's diligent record keeping provides us with useful insights into his use of the tent observatory during their second stop-over in Queen Charlotte Sound:

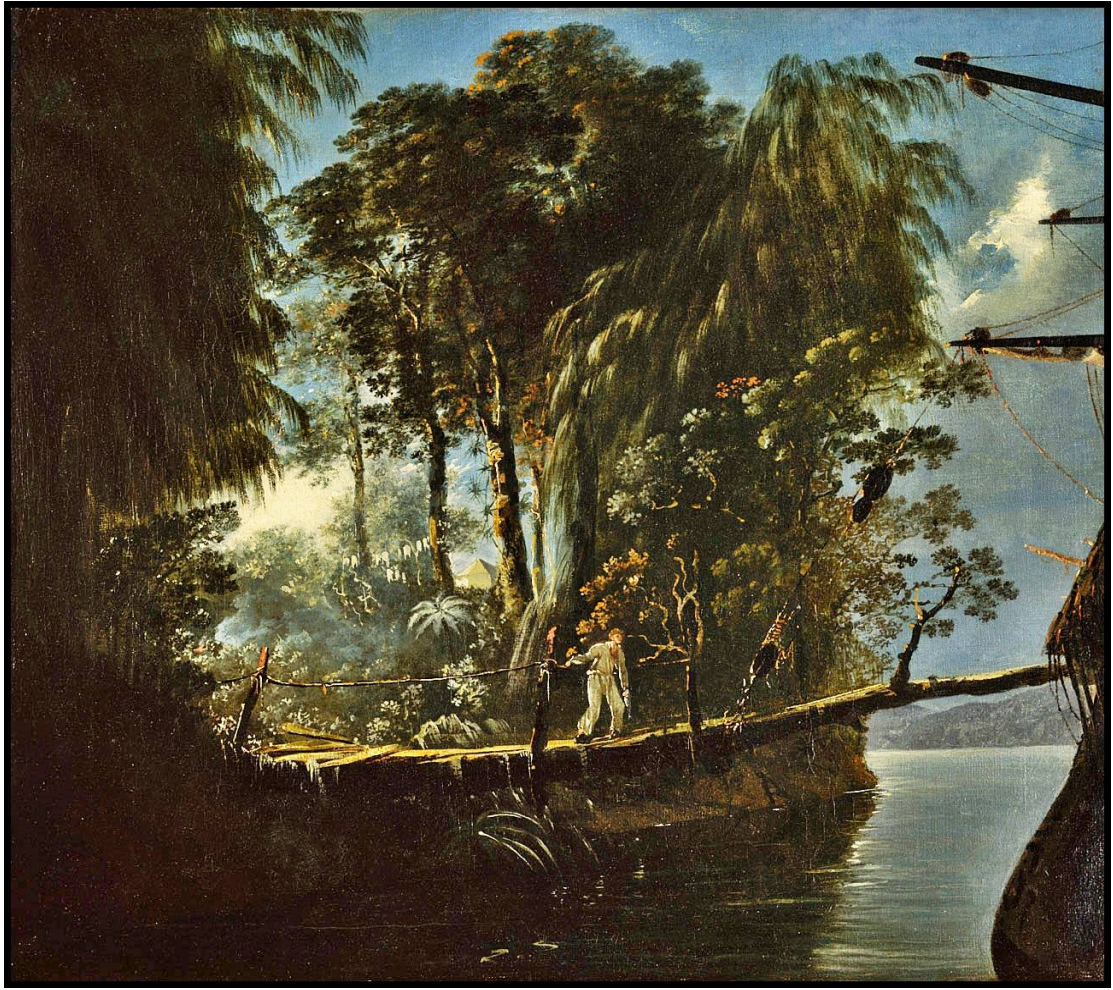

**Figure 3**: "View in Pickersgill Harbour, Dusky Bay, New Zealand"; artist: William Hodges. In a distant forest clearing a tent observatory has been pitched. (National Maritime Museum, Greenwich, Ministry of Defense Art Collection; low-resolution image free to use).

> [On 3 December 1773, I] Carried my Tent Observatory & Instruments on Shore & set all up by the side of the Garden a little distance from the watering place [in Ship Cove] ... The place where my Observatory stood when we were here before bearing E 5½° W per Compass distant ... (Bayly, 1773c).

Bayly set out to obtain a series of gravity observations with the cumbersome astronomical regulator. Unfortunately, however, during this stop-over, security became a serious concern, as we learn from his notes of 14 December (see also Henry and Marra, 1775; Bosloper, 2017):

> I was up late observing & having taken some altitudes of Stars to the East & having set my Alarem [*sic*] to call me, to take them to the West, I went to bed, having nailed my old great coat at the entrance of my tent, at the inside of which I always placed the outside case of my Astronomical Quadrant ... In this box I kept my Lumber, such as tools, nails, &c. ... After I had been in bed & had slept some time, I was awaked [*sic*] by the rattling or noise of the lid of the box. I jumped up in the bed & took my gun ... calling at the same time, who was there, but could neither hear nor see anything ... Soon after I searched for my hat but could not find it ... on going out of the Tent Obs.[y] I found it open & half the lid of the box at some dist. from the tent, & by feeling I found my hatchet & saw & hammer were gone out of the box. (Bayly, 1773a: 215–216).

Bayly's observatory design had proved is efficacy in storing and protecting the large astronomical regulators commonly carried on long-haul oceanic voyages. And so, on Cook's third voyage to the Pacific, tent observatories of the same type were deployed once again in Ship Cove, Queen Charlotte Sound, in February 1777 (e.g., Greenhill, 1970: 27; Orchiston,

2016). Cook, acting as the shipboard astronomer on H.M.S. *Resolution*, and Bayly, on H.M.S. *Discovery*, pitched their tent observatories close to one another in both Ship Cove (see Figure 4) and also in Nootka Sound at what is now Vancouver Island, Canada (Orchiston and Wells, 2020), between 29 March and 26 April 1778 (see Figure 5). Cook initially called their Nootka Sound mooring site Ship Cove, after his preferred site in New Zealand waters. It is now called Resolution Cove and is located at the southern end of Bligh Island.

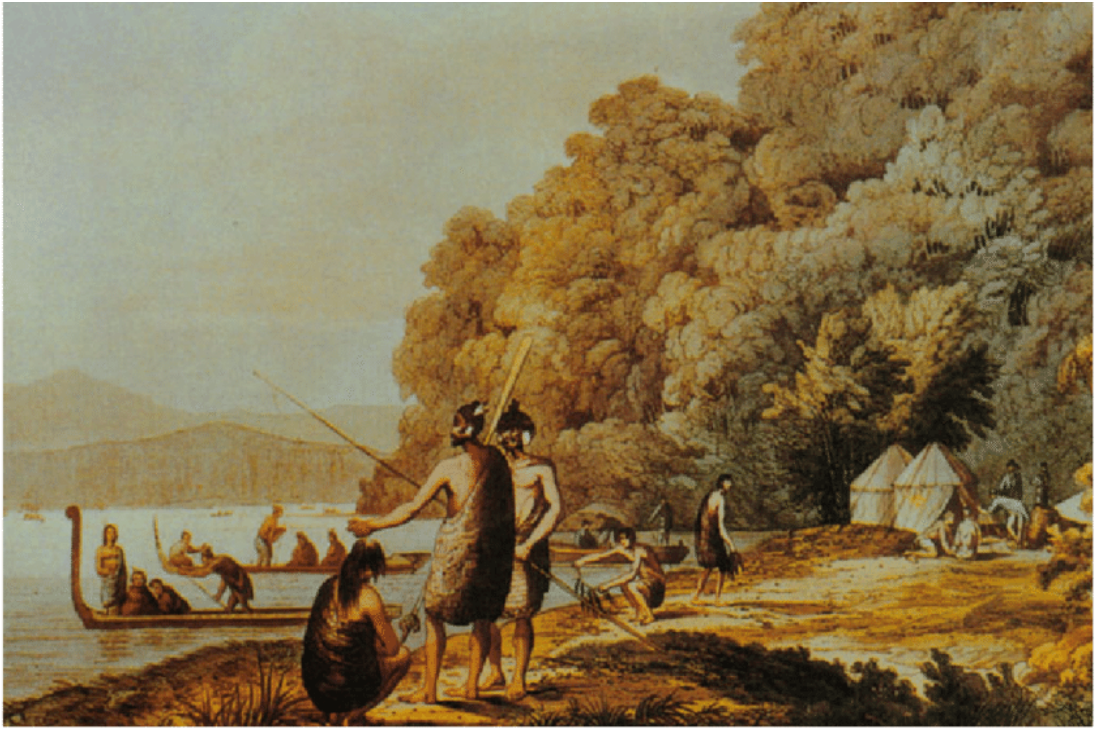

**Figure 4**: Tent observatories at Ship Cove in Queen Charlotte Sound, February 1777. Artist: John Webber (Courtesy: Suter Art Gallery, Nelson, New Zealand; out of copyright).

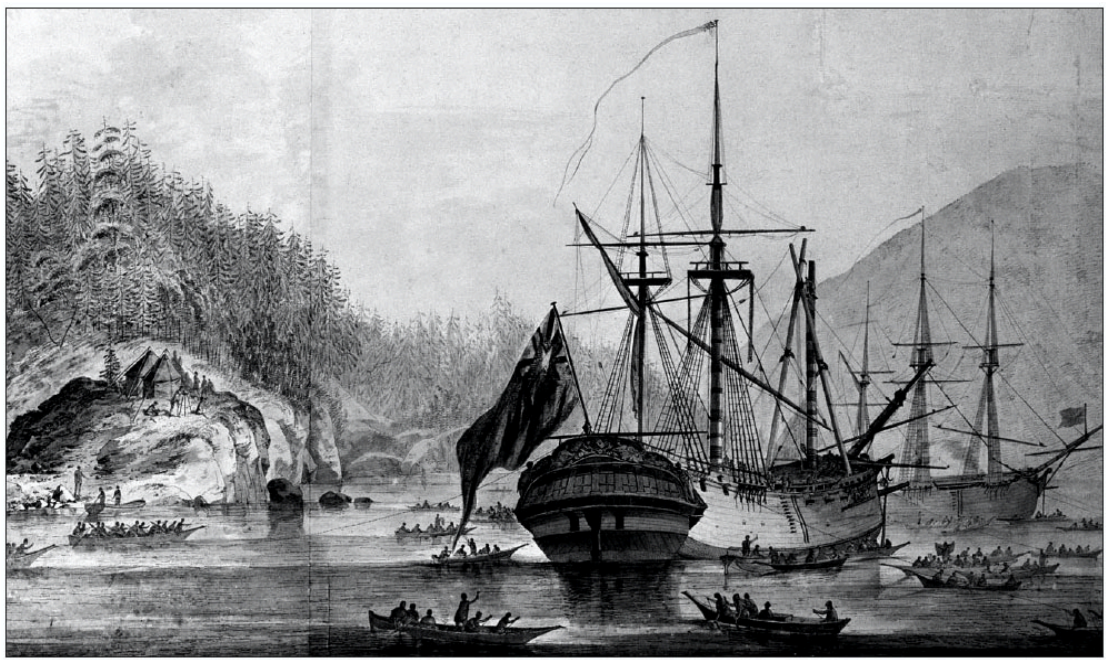

**Figure 5:** Temporary tent observatories are seen on the shore of Vancouver Island during a visit by Cook's ships *Resolution* and *Discovery* during Cook's third voyage to the Pacific. Artist: John Webber (1778). (National Maritime Museum, Greenwich, object No. PAJ2959; low-resolution image free to use).

## 3 STANDARD EQUIPMENT

Cook's three voyages to the Pacific served as practical trials of the latest tent observatory design. Satisfied that Bayly's design served the needs of long-haul expeditions well, tent observatories soon became standard equipment on voyages of exploration supported by the British Admiralty and the British Board of Longitude. By the late 1780s, Maskelyne—in his dual roles as Britain's Astronomer Royal and Commissioner of the Board of Longitude—proceeded to define a list of standard equipment for expeditionary voyages (Howse, 1979).

Standard observations Maskelyne expected his astronomers to undertake included determinations of latitudes and longitudes, magnetic variation and dip, air temperature, positions of headlands, islands and harbours, and the heights and times of tides (e.g., Beaglehole, 1999: 1500–1503). Maskelyne's instructions explicitly directed his astronomers to use the tent observatory they would be provided with:

> Observations to be made on shore:-
> 1. Whenever you land, if time permits, you are to set up the tent Observatory, and astronomical clock; either setting the pendulum exactly to the same length as it was of at Greenwich, before the voyage or noting the difference by the revolutions of the screw and divisions of the nut at bottom; ... (Maskelyne, 1801).

As a case in point, William Gooch (1770–1792), the astronomer on board the Vancouver expedition of 1791–1795, wrote to his parents (Gooch, 1791a; my emphasis),

> Perhaps you'll like to know what Instruments I'm to take abroad, … most of them the same that went with Cap.$^n$ Cook:
> 1. An Astronomical Clock
> 2. A Journeyman Clock
> 3. An Alarum Clock
> [4.] A Good Watch w$^{th}$ Second Hand
> [5.] An Achromatic Telescope of 46 Ins. Focus w$^{th}$ a divided Object-Glass Micrometer
> 6. A Reflecting Telescope
> 7. A Vertical Circle with an Azimuth-circle for taking altitudes and azimuths
> 8. A Transit Instrument of 4 Feet with a Level & upright wooden Posts
> 9. A Marine dipping needle [a freely suspended magnetic needle used to determine the local magnetic inclination]
> 10. A small Pocket compass
> 11. A Set of Magnetic Bars to change the Poles of the dipping Needle
> 12. A Burton's Theodolite with Stand
> 13. A Hadley's Sextant by Dollond
> 14. Another by Troughton
> 15. Two large Thermometers
> 16. Two Thermometers with wooden Scales, by Ramsden
> 17. A Portable Barometer by Burton
> 18. A Bason [basin] to hold Quicksilver [mercury] with Glass Roof
> 19. Quicksilver in a Bottle
> 20. A Night Telescope
> 21. A Steel Gunter Chain
> 22. A Knight Azimuth Compass
> **23. A Portable Tent Observatory**
> besides Books

Maskelyne had explicitly ordered that a portable observatory tent was made for Gooch to take on the Vancouver expedition (Maskeyne, 1791). By this time, however, it appears that tent observatories could now be supplied directly by the Navy Board, without the need to involve the Astronomer Royal (e.g., MacDonald and Withers, 2016: 31). Gooch's standardised set of equipment broadly resembles that taken by William Dawes (1762–1836), another of Maskelyne's protégés, on the 'First Fleet', the convoy of eleven ships that brought the first British convicts to Australia in 1787–1788 (de Grijs and Jacob 2021a; and references therein). During this 250-odd day voyage, however, Dawes was able to disembark his observatory tent only once, during their month-long stop-over in the port of St. Sebastian (Rio de Janeiro). Captain Arthur Phillip (1738–1814), commanding officer of the First Fleet's flagship H.M.S. *Sirius* and Governor-designate of the new colony in New South Wales, secured permission for the astronomer to establish a temporary observatory on Enchados island, allowing Dawes to install his instruments and perform the type of observations required by the Board of Longitude.

Upon the First Fleet's arrival in Sydney Cove, Dawes initially set up his tent observatory (Haynes et al., 1996: 31–32) on what is today called Dawes Point, although Dawes originally named the promontory Point Maskelyne, after his patron (de Grijs and Jacob, 2021b). Eventually, he established a wooden observatory. For that latter observatory's revolving white conical canvas roof, which included a retractable shutter, he most likely repurposed the tent observatory's fabric (e.g., Kerr, 1986).

## 4 ROUTINE OBSERVATIONS

### 4.1 Early French expeditions

During the same period that Dawes set out to establish the first permanent observatory in the New South Wales colony, a French expedition under Jean-François de Galaup, comte de Lapérouse (1741–1788) had temporarily settled on the northern shore of Botany Bay, within a day's trek of the British settlement at Sydney Cove. Among its 10 scientists, the Lapérouse expedition included the astronomer Joseph Lepaute Dagelet (1751–1788). Lapérouse's Royal instructions of 15 February 1785 included explicit orders to establish an observatory in any of the expedition's ports of call:

> Immediately upon arriving in a harbour, he will select an appropriate site on which to erect the tents and the observatory, and will set up a guard. … Separately from observations relating to the determination of latitudes and longitudes, for which every known and practicable method will be used, and those needed to assess the declination and inclination of the dipping needle, he will ensure that any celestial phenomenon which may be visible be observed; and on every occasion he will give the astronomers all the help and facilities necessary for the success of their work. (Dunmore, 1994: cxlii–cxliii).

And so, the French had built a stockade, an observatory and a garden for fresh produce on the Sydney headland now known as La Perouse. We learn from an entry of 2 February 1788 in the private journal of Lieutenant Philip Gidley King (1758–1808) that the temporary French settlement was already set up for astronomical observations to commence:

> After dinner I attended ye Commodore & other Officers onshore where I found him [Lapérouse] quite established, having thrown round his Tents a Stoccade, guarded by two small guns in which he is setting up two Long boats which he had in frames, an observatory tent was also fixed here, in which were an Astronomical Quadrant, Clocks &c under the Management of Monsieur Dagelet Astronomer, & one of y$^e$ Académie des Sciences at Paris. (King, 1788).

Shortly after the departure of the Lapérouse expedition from Botany Bay, both of their vessels, the *Boussole* and the *Astrolabe*, and all of the crew disappeared. Therefore, in September 1791, the French Assembly agreed to send a new expedition to Australian waters, in part to search for any trace of the ill-fated Lapérouse expedition. Antoine Raymond Joseph de Bruni, chevalier d'Entrecasteaux (1737–1793), was selected as commander of the expedition's main frigate, *Recherche*, which was accompanied by a second frigate, *Espérance*.

The d'Entrecasteaux expedition had been equipped with tent observatories for the convenience of the shipboard astronomers whose main duties involved determining accurate, land-based longitudes and latitudes using astronomical observations supported by marine timepieces. As such, establishing the tent observatories on shore was considered a priority upon arrival in any harbour where conditions would allow the astronomers to do so. We learn from the expedition's journals that on arrival in Recherche Bay, Tasmania, in April 1792 a tent to house the astronomical equipment was pitched behind the sandy beach on Observatory Point, now Bennetts Point (Mulvaney, 2007: 43–44). A specific aim of the d'Entrecasteaux' astronomers at this time was to observe a predicted eclipse of one of Jupiter's moons, which had been forecast to occur on 26 April 1792. Design particulars of the French tent observatories in use at the time are not available.

## 4.2 British achievements

Meanwhile, the British Admiralty had dispatched Captain George Vancouver (1757–1798) on a 4½ year-long circumnavigation of the globe. As we saw already, William Gooch had been assigned the astronomer's responsibilities, and he was directed to meet the Vancouver Expedition by joining Captain Richard Hergest (ca. 1754–1792) on the store ship *Daedalus*, bound for a rendezvous at Nootka Sound. For the first time on a British expedition, Gooch had been provided with different sets of instructions for shipboard and shorebound astronomical observations (Board of Longitude, 1791; see also Beaglehole, 1999). The large and cumbersome transit instrument and the astronomical regulator were explicitly meant for use on land only. These instruments required repeated observations for calibration purposes, before they could be employed to obtain useful new data.

However, opportunities for shorebound observations were few and far between, and they were ultimately at the discretion of the commanding officer. Gooch realised this early on, stating that "I intend taking proper instruments on shore to measure the perpendicular height of the Pike of Teneriffe if I can persuade Hergest to stay long enough" (cited by Dunn, 2014: 89). His best chance to obtain an extended series of astronomical observations came when the *Daedalus* stayed in the harbour of St. Sebastian for an extended sojourn to allow for repairs and re-provisioning. Gooch established a tent observatory for his astronomical regulator, the larger telescopes and other, mostly larger instruments. His tent observatory was almost identical to those used on Cook's second and third voyages, at "… 10 Feet [305 cm] Diameter, made in 12 Frames, with a Moveable Roof on Kirbs, made up in 3 thickness's, Door & proper fastnings fix[d] on a Bottom folding Kerb, & pack'd up in Cases" (Smith, 1791; see also Davies 1994). It had been purchased for £20 from Nathan Smith, proprietor of the Patent Floor Cloth Manufactory in Knightsbridge, London (Dunn, 2014: 89).

Gooch was all set to commence his astronomical observations in the harbour of Rio de Janeiro by 6 November 1791. He had been allocated a piece of land for his observatory for which he paid a local widow 2 shillings a day in rent (Gooch, 1791b). However, the tropical Brazilian weather did not cooperate. On 15 November, the prevailing strong winds prevented him from opening the observatory's roof, "… as it was likely the wind would have carried it away …" (Gooch, 1791c). In compliance with his official instructions, he had successfully installed his Ramsden's universal theodolite, but he seems to have been unable to deploy it properly, which did not help to brighten his mood:

> The Observatory & Instruments were taken down yesterday [23 November 1791] & carried into a house on the Island where the Observatory was put up. & this morning they were brought on Board. - It is rather a discouraging thing to me (at first setting out) to meet with such unfavourable weather. The Observatory was up 10 days before I could do any thin the weather being constantly cloudy. (Gooch, 1791d).

Other than being affected by the vagaries of the weather, his observatory's location near a well-populated area came with additional drawbacks in the form of almost continuous visits by the local gentry:

> The Sunday after the Observatory was put up, and the Instruments fixt in it, several Gentlemen of Rio-Janeiro with some Ladies, came over to see the Observatory ... One of the Gentlemen spoke a little Latin, and English both, so that by one means of other, we could understand a great many things, - but somehow, he strangely mistook me in one thing. – I ask'd him to propose to the Ladies, their looking at a distant Object through a Telescope. – He look'd me very hard in the face, and ask'd me with surprise, what it was I said. – I repeated the same thing. – When he <u>sternly</u> replied <u>no</u>: - adding that the Ladies present, were people of Rank and Credit, - and that he was amaz'd at my Proposal. – I saw immediately, he had misunderstood me, and could not help laughing, at so strange a mistake; - however, I soon explain'd myself, and then he laugh'd as well as I, shook me by the Hand, and apologiz'd for his Misconstruction of my Words, - but what that Misconstruction literally was, I never knew; - however the purport of it is easily perceiv'd from his answer. (Gooch, 1791e).

> I had been up all night in the Observatory, and was oblig'd to leave my visitors to go a few minutes in the afternoon to the Observatory to take some Altitudes of the Sun; - after that, when I had sat with them some time, I grew very sleepy on a sudden, so much so, that I was oblig'd to tell Dobson, to explain to them, that I had sat up all night, and was retir's to lay down to sleep for half an hour; - but when they were about to wake me as order'd, my

Visitors desir'd I might not be awake, as they were sure (they said) I wanted rest, - so I wasn't present when they went away. – But when I awoke some time afterward, I turn'd out to undress, and told Pitts to set the Alarum, to the time of the Moon's passing the Meridian, - When the Alarum ran down I look'd out and saw 'twas cloudy, so I went to Bed again and slep't 'till nine in the Morning. (Gooch, 1791f).

We already saw that the Flinders expedition of 1801–1803 also routinely employed tent observatories. On the expedition's initial leg from England to Table Bay, they benefited from the services of John Crosley (1762–1817), who had been assigned to the voyage as shipboard astronomer. Crosley stayed with the expedition until their arrival at the Cape of Good Hope. He did not continue on the next leg to Australia on account of his deteriorating health. He was seriously unwell with gout and requested to be discharged when they reached the Cape.

On 18 October 1801, Crosley was accompanied by Matthew and Samuel Ward Flinders on a shore excursion to look for a suitable location to set up the expedition's tent observatory and marquee. This was their final opportunity to check and calibrate their clocks and instruments for longitude determination before setting off on the Australian survey. They settled on a location near the Dutch East India Company's garden in Cape Town, where work on the observatory continued until 1 November, duly guarded by a complement of Marines (Morgan, 2016). Among the clocks included in the astronomer's arsenal, the Earnshaw astronomical clock, No. 543, was equipped with a weight-powered regulator. It was used on shore to calibrate the expedition's other clocks against the stars (Davies, 1997).

Before his departure back to England, Crosley provided detailed instructions to Matthew and Samuel Ward Flinders as regards the adjustments they were to make to the clocks and the universal theodolite (Crosley, 1801). Crosley and Matthew Flinders then took their time to ensure that all clocks were properly calibrated for use on the expedition's next leg (Morgan, 2017), while also endeavouring to establish Cape Town as a geographic anchor location for future expeditions. However, as Flinders would later confess, the observatory tent's small size was "… a great obstruction to our operation …" on shore (Flinders, 1802).

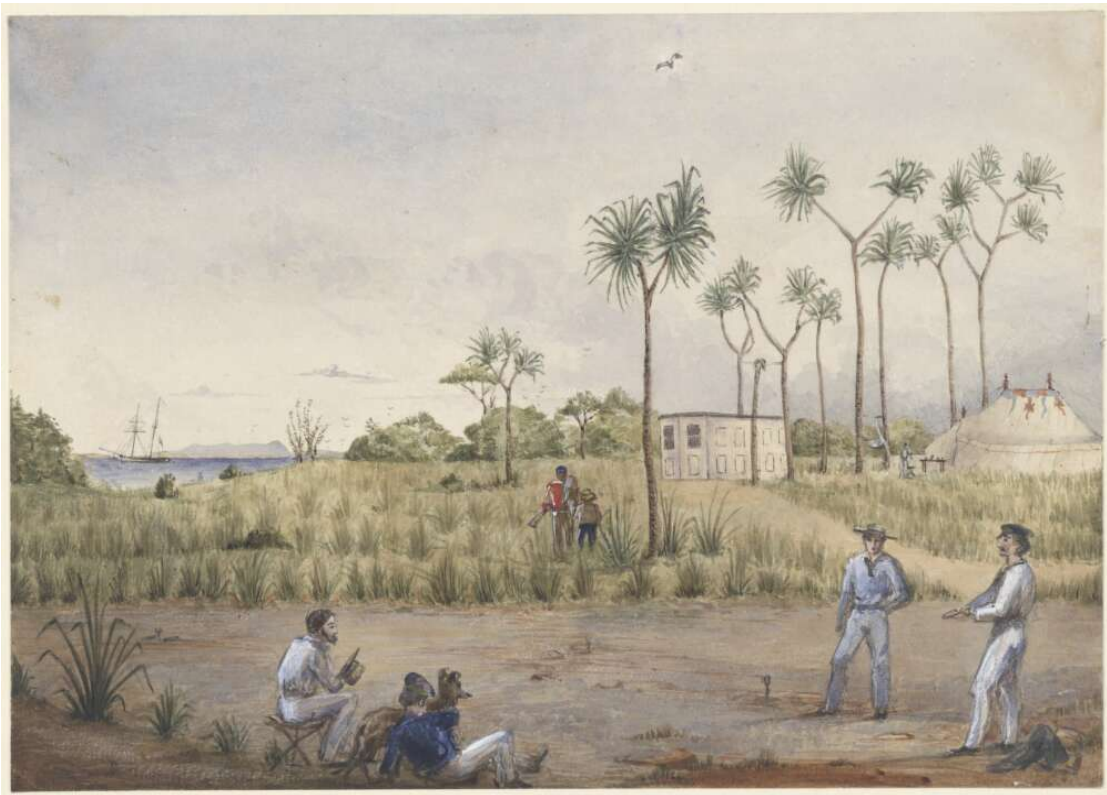

**Figure 6**: Portable observatory at Cape Upstart, Queensland, April 1843. Artist: Edwin Augustus Porcher. (National Library of Australia, PIC Drawer 3521 #R5706; out of copyright).

The Flinders brothers had their first opportunity to engage in extensive land-based astronomical observations upon arrival at King George Sound, near the entrance to the deep-water harbour of present-day Albany, Western Australia. Following their arrival in early December 1801, they spent nearly a month *in situ*, which allowed them to pitch the observatory tent and install most of their instrumentation on shore. Their initial data set consisted of a series of 31 lunar distance measurements for comparison with the longitudes implied by their marine chronometers. This combination of astronomical observations and clock-rate measurements allowed them measure to the daily rates of the timekeepers and to record any deviations from apparent Greenwich mean time. Matthew, but increasingly Samuel Ward Flinders, continued to obtain careful astronomical observations, including equal altitudes of the Sun at various locations along Australia's south coast—at Lucky Bay and Goose Island Bay, in the Recherche Archipelago, off the southern coast of present-day Western Australia; at the head of the Great Australian Bight; near Port Lincoln, at the head of Spencer Gulf and at Kangaroo Island, South Australia; and at Port Phillip (off present-day Melbourne, Victoria; Morgan, 2015).

**4.3 Changing priorities**

By the end of the eighteenth century, compact, accurate marine chronometers were becoming increasingly available. The need for shipboard astronomers, tent observatories and cumbersome astronomical regulators on long-distance oceanic voyages of exploration thus gradually decreased. For several decades, such voyages would continue to routinely take along the traditional complement of astronomical instrumentation, but their navigators increasingly relied on the novel maritime timepieces. Figure 6 is a representation of a tent observatory at Cape Upstart (south of present-day Townsville, Queensland) in April 1843, during an expedition to map northeastern Australia, including the Great Barrier Reef, from H.M.S. *Fly*, under Captain Francis Price Blackwood's (1809–1854) command.

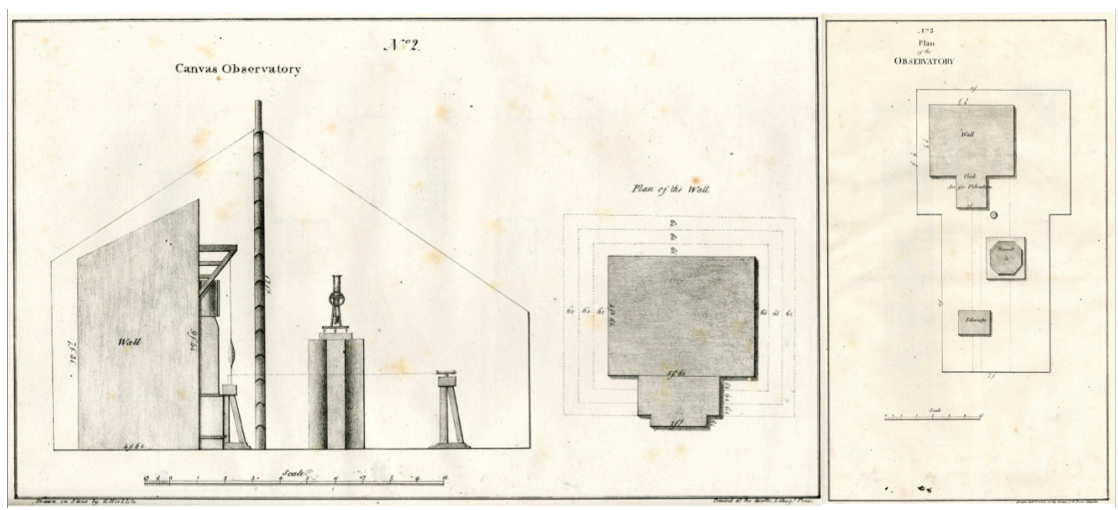

**Figure 7**: Canvas observatory used on an expedition led by John Goldingham, F.R.S., from Madras to Sumatra in 1824. (Goldingham, 1827: Plates No. 2 and 5; Courtesy: Institute of Astronomy, University of Cambridge; Public Domain Mark 1.0, Attribution 4.0 International licence).

Tent observatories and shorebound heavy instruments did not become surplus to requirements for some time, however, but their use gradually shifted to applications where temporary observatories were needed owing to the occurrence of a specific celestial event at a predetermined time. We already encountered one such example, that is, the desire of the d'Entrecasteaux expedition's astronomers to observe a specific eclipse of one of Jupiter's moons on 26 April 1792, when the expedition was moored in Recherche Bay, Tasmania (although on that occasion it took longer than expected to pitch the tent and set up the instruments, so that the event was eventually not observed). Other examples include the use of a tent observatory on an expedition led by John Goldingham (ca. 1766–1849) from the British observatory in Madras (Chennai), India, to Sumatra, Indonesia (Goldingham, 1824: Plate No. 2; reproduced as Figure 7, left). Similarly, in 1882 Admiral Pelham Aldrich (1844–

1930), commander of survey vessel H.M.S. *Fawn*, and his astronomer, Stephen Joseph Perry (1833–1889), observed a Venus transit from an improvised tent observatory in Madagascar (Hingley, n.d.; see Figure 8).

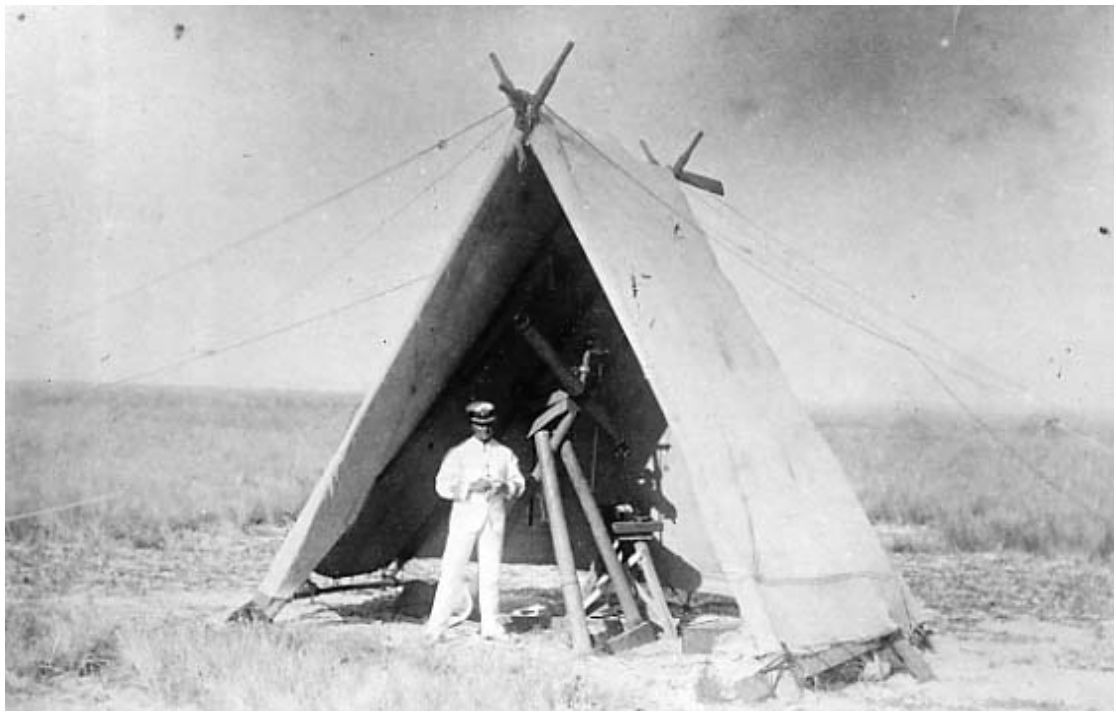

**Figure 8**: Commander Pelham Aldrich of H.M.S. *Fawn* prepares to observe the 1882 Venus transit from an improvised tent observatory in Madagascar. (Royal Astronomical Society. Reference: RAS ADD MS 209, p. 21, middle; reproduced with permission).

In preparation for the total solar eclipse of 17 August 1868, Major James Francis Tennant (1829–1915) and Dr. Edmund Weiss (1837–1917) advocated for a scientific expedition to the Indian subcontinent, to be supported by the Royal Astronomical Society (Tennant, 1867a, 1867b; Weiss, 1867). Major Tennant travelled to India to record the event. Upon arrival in Calcutta (Kolkata), Tennant decided to establish a base in Guntoor (Guntur) in Andhra Pradesh, a city accessible by both river and road. On 15 June 1868, Tennant and the photographers in his party left Calcutta by steamer down the coast, along with "… all our apparatus" (Tennant, 1869: 6), eventually anchoring at Coconada (Kakinada), on the Andhra Pradesh coast, on 22 June. Their equipment was transferred to Bez(a)wada (Vijayawada) by barge, with the expedition members following on a smaller steamer. They travelled the final 30 km by road, arriving in Guntoor on 3 July 1868 (Orchiston et al., 2017: Ch. 25.2).

The grounds of the home of the Subcollector of Guntoor, one Mr. H. H. Wilson, proved to be an ideal observatory site (see Figure 9). Their first task was to set up the Royal Astronomical Society's Sheepshanks' Telescope (which had been shipped from England) on a concrete pillar in a tent observatory that had been purpose-designed in Bezawada:

> The iron pillar of the Equatorial-stand rested on a stone 2ft. 6in. [76.2 cm] in diameter, forming the top of a brick pillar of the same diameter, and 4 feet [121.9 cm] deep, standing on compact gravel. The surface of the stone was level with that of the ground, and the whole instrument was enclosed in an octagonal tent, 10 feet [305 cm] in diameter, with a pyramidal roof … When the instrument was not in use, a waterproof sheet protected it from leakage which, in heavy rain, is almost unavoidable in a shelter of this sort. (Tennant, 1869: 15–16).

A separate tent observatory, also manufactured in Bezawada and delivered to Guntoor on 20 July 1868, was dedicated to house the expedition's Browning–With reflector telescope. The latter was installed on

... a solid brick foundation capped by stone, precisely similar to what has been described for the Sheepshanks' telescope, save that to allow the clock-weight to pass down, a cast-iron pipe, 12 inches [30.5 cm] in diameter, and closed at the bottom, was sunk, and of course, its upper end partly imbedded in the pillar. The tent ... was 12 feet [366 cm] in diameter ... (Tennant 1869: 29-30).

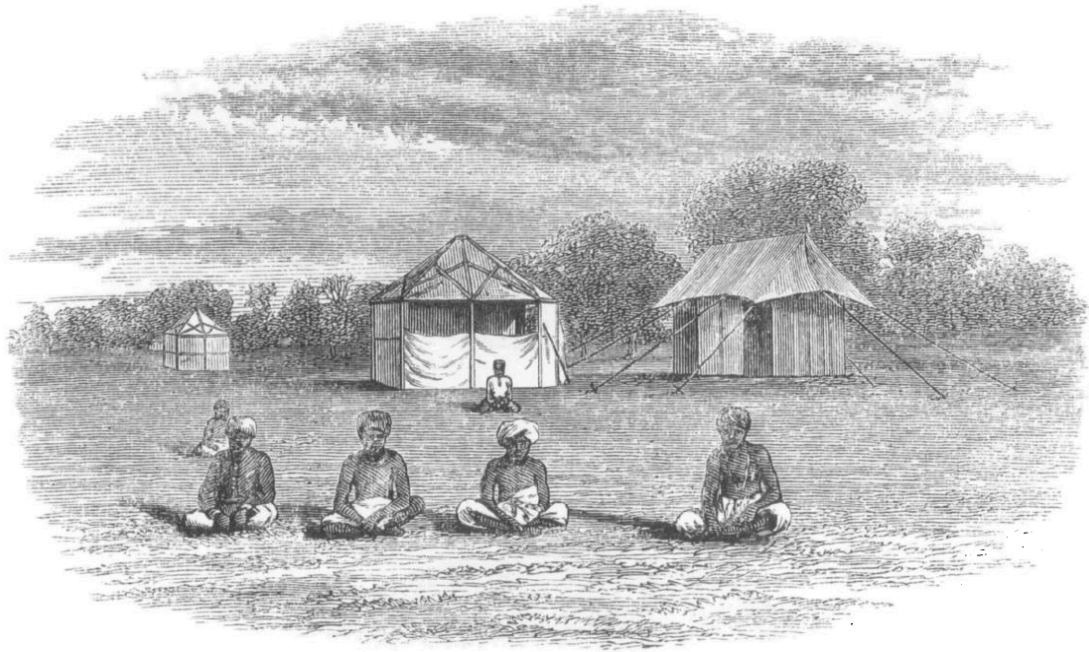

Sheepshanks.       Silver-glass.       Dark Tent.

**Figure 9**: "A View by Sergeant Phillips, R. E. [Royal Engineers], showing Observatory Tents" at Guntoor. (Tennant, 1869: 49; out of copyright).

## 5 FOCUS SHIFT

By the middle of the nineteenth century, the era of shipboard astronomers carrying tent observatories on long-haul voyages of exploration was all but over. However, the concept of the astronomical tent observatory remained attractive to a certain class of explorers, specifically those undertaking difficult treks to remote areas. Work done from those overland expeditions often involved both meteorological and astronomical observations.

The idea of using portable observatories on overland expeditions was not new. It had been embraced as early as the Venus transit of 1769 (which had sent Cook halfway around the world to Tahiti and beyond). To also observe the 1769 Venus transit closer to home, the Royal Society of London and the British Board of Longitude had selected John Bradley (1728–1794), a nephew of Britain's third Astronomer Royal, James Bradley (1692–1762), to lead an overland expedition to the Lizard peninsula in Cornwall, England. He was given a portable tent observatory to protect the expedition's equipment. Although their main aim was the observation of that year's transit of Venus, the Board of Longitude was also particularly interested in establishing accurate longitudes and latitudes as a shipping aid along the hazardous coastal waters around the Cornish mainland.

With another Venus transit event off the cards for another 105 years, the British scientific and maritime institutions had pooled their resources in an effort to cover the event from as many vantage points around the world as feasible. And whereas Bradley did not have to subject himself and his crew to the hardships of months at sea, the overland journey to the tip of Cornwall was still stressful and difficult (Baker, 2012). His letters to Maskelyne offer clear testimony of his great anxiety about reaching Lizard Point, setting up the tent observatory and then carrying out the requisite observations as illness wracked his body and foul weather threw a spanner in the works (Bradley, 1769: 4–26). Road access was poor, yet they chose to journey by wagon, "… often very uneasy for fear of the Carriage being overturn'd". Meanwhile the prevailing weather conditions in May 1769 were "… so bad we have seen neither Sun or Horizon with any distinctness" (Bradley, 1769: 7), which in turn caused them significant difficulties in erecting the tent observatory.

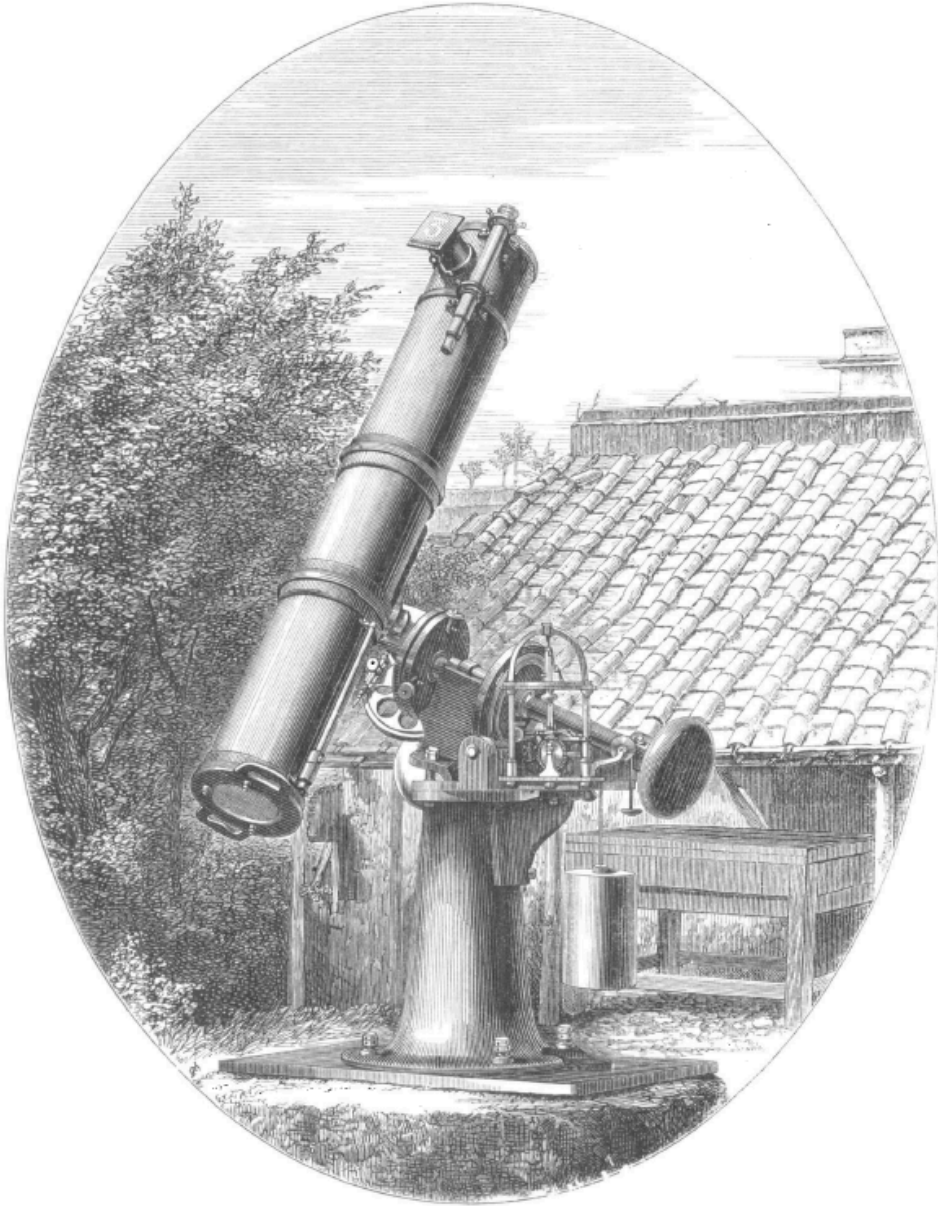

**Figure 10**: "Copy of a Photograph of the Silver-glass Equatoreal" [*sic*] used for solar eclipse observations at Guntoor. (Tennant, 1869: 51; out of copyright).

Land-based surveys using tent observatories were proceeding apace at the height of the era of shipboard tent observatory use. The Great Trigonometrical Survey of India started in 1802 and continued for more than 70 years. Although permanent observatories were eventually established to service the Survey's theodolites and rate the box chronometers, they replaced the tent observatories that had initially been used to establish the lay of the land (Davies, 2022). Halfway around the world, in 1803 United States' President Thomas Jefferson (1743–1826) had purchased the Louisiana territory from France, which nearly doubled the area of the new United States. Jefferson commissioned a great land survey (1804–1806) to map the extent of the new country, commanded by Meriwether Lewis (1774–1809) and William Clarke (1770–1838). This expedition, also known as the Corps of Discovery, initially had access to only one chronometer. The surveyors planned to determine the longitudes of the locations they passed through using the lunar distance method, employing an equatorial telescope (which compensates for the Earth's rotation; for an example, see Figure 10) sited in an observatory tent. However, Lewis decided to bring along his own Arnold chronometer to more efficiently facilitate these geographic position determinations (Davies, 2022).

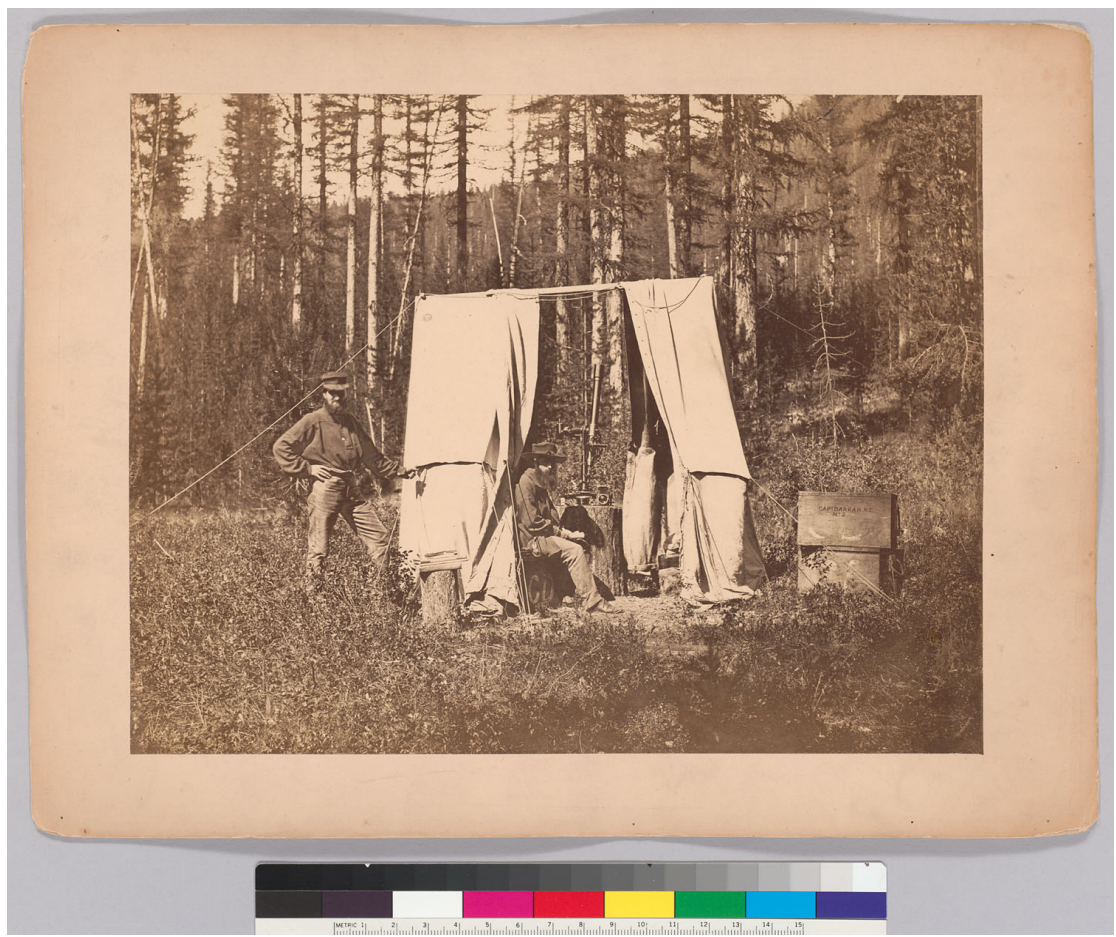

**Figure 11**: Observatory tent, North American Boundary Commission (1861). The photograph shows Captain Charles John Darrah (1835–1871), Royal Engineer, seated in the tent, and Captain Robert Wolseley Haig (1830–1872) of the British North American Boundary Commission posed at an astronomical observation tent with theodolite, during the marking of the boundary line between Canada and the United States. The box at right is marked "Capt. Darrah, R.E., No. 2". (Library of Congress Prints and Photographs Division, USA. Reproduction No.: LC-DIG-ppmsca-03143; no known restrictions on publication).

Tent observatories in support of astronomical observations to determine longitudes and latitudes were in continuous use in the United States for several decades thence. Employed by the British North American Boundary Commission, Samuel Anderson (1839–1881), an English surveyor, astronomer, cartographer and Royal Engineer, led the way in mapping the Canada–United States border in the 1860s and 1870s. They carried top-of-the-line box chronometers that were religiously wound and carefully checked (Rees, 2008). Anderson and his scouts selected and marked the best sites for the 40 astronomical stations they were meant to establish (see Figure 11), and they provided initial estimates of the sites' longitudes. British and American teams of astronomers would follow Anderson's vanguard and establish these stations between the Red River and the Continental Divide:

> On approaching the site selected for an astronomical station, usually at about 3 pm., though sometimes much later, the first step was to select, for the observatory tent, an elevated spot from which an uninterrupted sight line could be obtained to a distance of about three-fourths of a mile [1.2 km], either due north or due south. The camp was then pitched at a short distance off, so that neither the north or south not the east or west lines from the observatory tent came within 100 yards [91 m] of it. (Featherstonhaugh, 1876).

This same procedure was followed in 1872 across the Continental Divide to the Pacific Ocean so as to determine the 49th northern parallel (Rees, 2008). As pointed out by Albany Featherstonhaugh (1840–1902), Royal Engineer, in the quotation above, it was crucial that the explorers' camp was located some distance from the sensitive astronomical and surveyors' instruments; vibrations from horses, wagons and even men walking past could easily ruin a set of observations.

By the end of the nineteenth century, astronomical tent observatories had become largely obsolete. The concept of 'tent observatories' survived, however, but no longer as a means to determine one's longitude, observe stellar transits or record specific celestial events. Tent observatories became associated with meteorological observations in remote and harsh conditions (e.g., *The Daily Telegraph*, 15 December 1897; Ingleby, 1897–1902; Davies, 2014; Higgins, 2019), and they were also used for geodetic observations. The major difference between astronomical and meteorological or geodetic tent observatory designs was found in the construction of their roofs. Astronomical tent observatories were usually equipped with retractable or folding roofs, a design feature that was not required for meteorological or for most geodetic observations.

The development of more compact, more accurate, more precise and more stable chronometers between the late eighteenth and the late nineteenth centuries signaled an end of sorts for the astronomical tent observatory. In fact, the need for tent observatories continued into the twentieth century, specifically for geodetic astronomical observations. As late as the 1980s, such expeditions[3] continued to employ purpose-designed canvas tents to protect their delicate theodolites from both the elements and external disturbances, and to provide the observing team with shelter. When satellite-based technologies eventually negated the need for this type of observation, the astronomical tent observatory finally met its end. Today, demand from amateur astronomers, from novices to semi-professionals, ensures continued development of modern, versatile and state-of-the-art observatory tents (e.g., Allison, 2022; Pearson, 2022).

**6 NOTES**

[1] Another example involves a voyage back from Sydney, via China, to England in 1802, where James Inman (1776–1859) paid £8 as "A Present to [a] Convict Servant on leaving Port Jackson [Sydney Harbour], for his great Care, & Attention in Watching at the Tent Observatory, when Indisposition prevented me from sleeping at it myself". (Inman, 1802).

[2] Richard Pickersgill (1749–1779) was the *Resolution*'s third lieutenant.

[3] In Australia, such expeditions were undertaken by the Royal Australian Survey Corps, for instance. For examples of the type of canvas tents used on their expeditions, see e.g. Wise (2021–2022) or https://www.xnatmap.org/adnm/ops/prog/rafgeosvy/15.htm.